\documentclass[%
 reprint,
 amsmath,amssymb,
 aps,
]{revtex4-1}

\usepackage[utf8]{inputenc}
\usepackage[T1]{fontenc}
\usepackage{graphicx}% Include figure files
\usepackage{dcolumn}% Align table columns on decimal point
\usepackage{bm}% bold math
\DeclareFontFamily{OT1}{pzc}{}
\DeclareFontShape{OT1}{pzc}{m}{it}{<-> s * [1.15] pzcmi7t}{} % CONTROLA O TAMANHO DA LETRA
\DeclareMathAlphabet{\mathpzc}{OT1}{pzc}{m}{it}
\begin{document}

\preprint{APS/123-QED}

\title{Exploring the Thermodynamics of Confining Models}% Force line breaks with \\
\thanks{Poster presented at XIV Hadron Physics, $18^{th} - 23^{rd}$ March, 2018 in Florianopolis (Brazil)}%

\author{A. V. Silva}
 \email{apollovitalinosilva@gmail.com}%Lines break automatically or can be forced with \\
\author{B. W. Mintz}%
 \email{brunomintz@gmail.com}
\affiliation{%
Departamento de Física Teórica, Univerdade do Estado de Rio de Janeiro - UERJ, Rua São Francisco Xavier 524, 20550-013, Maracanã, Rio de Janeiro, RJ, Brazil\\
}%


%\affiliation{
%\affiliation{
%\affiliation{%



\begin{abstract}
Establishing a description for confinement is not something simple. In order to try to understand a little about this phenomenon, we will explore the thermodynamics of models that try to describe it in terms of propagators with violation of positivity. In this work, ``confinement'' is always understood in the sense of positivity violation of the propagator of the elementary fields. For simplicity, we will define a model for scalar fields with a momentum dependent and nonlocal mass term. One of our objectives is to verify the thermodynamic properties of this Lagrangian in order to analyze possible inconsistencies. For this we use the functional formalism of Quantum Field Theory  at finite temperature, from which we obtain the partition function and, consequently, the thermodynamic variables such as pressure, energy density, entropy density, etc. Then, we obtain the two-point function at finite temperature of the scalar field, in order to study whether or not there is a restoration of positivity (hence, deconfinement, in our language). 
\begin{description}
\item[Keywords]: Quantum Field Theory, Quantum Field Theory at finite Temperature, Model of Confining Fields, Gribov-Zwanziger Theory
\end{description}
\end{abstract}

\pacs{Valid PACS appear here}% PACS, the Physics and Astronomy
\maketitle

%\tableofcontents

\section{\label{sec:level1}INTRODUCTION}

Although the Higgs mechanism is considered to be the main responsible for the mass of leptons (such as the electron, for example) practically the whole mass of the visible matter of the Universe (not counting Dark Matter and Dark Energy) is concentrated in the atomic nuclei. The atomic nuclei are formed by protons and neutrons, which in turn are formed by three quarks each. At high energies, when the strong nuclear interaction is less relevant, we can infer what is the portion of the mass of the quarks that is due only to the electroweak interaction. The value obtained for the sum of the individual masses of the three quarks that form a proton corresponds to less than 2\% of the total mass of the proton. Thus, about 98\% of the mass of a proton or a neutron (and hence visible matter) originates from strong nuclear interaction.
%Embora o mecanismo de Higgs seja considerado como o principal responsável pela massa dos léptons (como o elétron, por exemplo), praticamente toda a massa da matéria visível do Universo (sem contar a Matéria Escura e a Energia Escura) está
%concentrada nos núcleos atômicos. Os núcleos atômicos são formados por prótons e nêutrons que, por sua vez, são formados por três quarks cada um. Em altas energias, quando a interação nuclear forte é menos relevante, podemos inferir qual é a parcela da massa dos quarks que se deve apenas à interação eletrofraca. O valor obtido para a soma das massas individuais dos três quarks que formam um próton corresponde a menos de 2\% da massa total do próton. Assim, cerca de 98\% da massa de um próton ou de um nêutron (e, portanto, da matéria visível) têm origem na interação nuclear forte.
 
The role played by the strong interaction is evident, but its detailed mechanism is little understood. The fundamental theory of strong interactions, Quantum Chromodynamics (QCD), can be solved by perturbative methods reliably only in very high energies. QCD is a gauge theory in which quarks (which are fermions) interact with each other through gauge bosons, called gluons. Because it is a non-abelian gauge theory, the gluons interact with each other, bringing enormous richness and complexity to the theory.
%O papel exercido pela interação forte é evidente, mas o seu mecanismo detalhado é pouco compreendido. A teoria fundamental das interações fortes, a Cromodinâmica Quântica (QCD), pode ser resolvida por métodos perturbativos de forma confiável apenas em altíssimas energias. A QCD é uma teoria de calibre na qual os quarks (que são férmions) interagem entre si através de bósons de calibre, chamados gluons. Por tratar-se de uma teoria de calibre não-abeliana, os gluons interagem entre si, trazendo enorme riqueza e complexidade à teoria.

One of the fundamental properties of strongly interacting matter is confinement, that is, there are no asymptotic states of a single quark, nor of a single gluon. At great distances (low energies), the interaction between quarks and gluons becomes so intense that it is not possible to isolate one of these particles from the others. Another way of understanding confinement is that strong interaction causes any state of a particle with non-zero color charge to be a non-physical state (for example, in violation of positivity). This second approach has been used as a confinement criterion for many years in analyses of the dynamic mass generation of quarks and gluons \cite{5Cornwall:1980zw,6Brambilla:2014jmp,7Alkofer:2000wg,8Aguilar:2004sw,8Aguilar:2008xm,9Maas:2011se}, for example. Such a violation (which is associated with a function, called the spectral function, be negative) can be roughly understood as a negative probability of propagation, and thus the absence of the quark state of the physical spectrum. Propagators that exhibit this violation, in momentum space, may have complex poles. This is how the model describes the confinement of an isolated quark. The nonlocal interaction that causes this effect can be understood as a result of the action of a background of gluons, on which the quarks move.

It is expected that, at high energies, quarks and gluons deconfinement occurs due to the phenomenon of asymptotic freedom. Confinement is an eminently non-perturbative phenomenon. Therefore, non-perturbative approaches, such as Functional Renormalization Group \cite{10Berges:2000ew,11blaizotl}, Dyson-Schwinger Equations \cite{12Cloet:2013jya}, Monte Carlo simulations in the lattice \cite{13Karsch:2009zz}, holographic methods \cite{14Brodsky:2014yha}, and finally effective models \cite{3Nambu:1961tp,15Klevansky:1992qe,16levy,16Pisarski:1983ms,16Rosenzweig:1979ay,16Lenaghan:2000ey,16Scavenius:2000qd,16Roder:2003uz}, are valuable tools in the study of strong interactions.
%Espera-se que, em altas energias, ocorra o desconfinamento de quarks e gluons, devido ao fenômeno da liberdade assintótica. O confinamento é um fenômeno eminentemente não-perturbativo. Portanto, abordagens não-perturbativas, como aquela do Grupo de Renormalização Funcional \cite{10Berges:2000ew,11blaizotl}, das Equações de Dyson-Schwinger \cite{12Cloet:2013jya}, das simulações de Monte Carlo na rede \cite{13Karsch:2009zz}, de métodos holográficos \cite{14Brodsky:2014yha} e, finalmente, de modelos efetivos \cite{3Nambu:1961tp,15Klevansky:1992qe,16levy,16Pisarski:1983ms,16Rosenzweig:1979ay,16Lenaghan:2000ey,16Scavenius:2000qd,16Roder:2003uz}, são preciosas ferramentas no estudo das interações fortes.

From the theoretical point of view, the exploration of physical models in a medium can also reveal important properties of the physical systems, as well as possible inconsistencies of the model that could not be very easily perceived in vacuum analyzes. Thus, this work fits as part of the great theoretical effort of High Energy Theoretical Physics in search of a better understanding of the strong nuclear interactions in extreme regimes. Such an undertaking requires methods that go beyond perturbation theory and have the capacity to predict experimental results, as well as the ability to compare with other theoretical methods. This is the case of the effective models method, which we will employ in our investigation.
%Do ponto de vista teórico, a exploração de modelos físicos em um meio pode também revelar propriedades importantes dos sistemas físicos, assim como possíveis inconsistências do modelo que poderiam não ser muito facilmente percebidas em análises no vácuo. Assim, este trabalho enquadra-se como parte do grande esforço teórico da Física Teórica de Altas Energias em busca de uma melhor compreensão das interações nucleares fortes em regimes extremos. Tal empreendimento exige métodos que consigam ir além da teoria de perturbação e que possuam capacidade de predição de resultados experimentais, assim como capacidade de comparação com outros métodos teóricos. Este é o caso do método dos modelos efetivos, que empregaremos em nossa investigação.

Finally, using the effective models method, we will propose a Lagrangian for a scalar field with a momentum dependent and nonlocal mass term. First, we will make an approach to the free scalar field, obtaining its propagator and the free partition function. Having the partition function in hand, we will then explore the thermodynamics of the theory, analyzing the behavior of some thermodynamic variables, such as pressure and entropy density.
%Por fim, utilizando o método dos modelos efetivos, iremos propor uma Lagrangiana para um campo escalar com um termo de massa não local, dependente do momento. Primeiramente, iremos fazer uma abordagem para ocampo escalar livre, obtendo seu propagador e a função de partição livre. Tendo a função de partição em mãos, iremos então explorar a termodinâmica da teoria, analisando o comportamento, de algumas variáveis termodinâmicas: pressão e densidade de entropia. 

%Within the general context of exploring effective non-perturbative models of strongly interacting matter under extreme conditions of temperature and density, we will investigate the thermodynamics of simple finite-temperature field theory models that describe confinement in terms of positively violating propagators. This violation comes from the fact that the spectral function is negative (remembering that we can rewrite the propagator in terms of it). As this is an eminently non-perturbative phenomenon, and that it is a property typical of non-abelian gauge theories, the exploration of toy models for confinement can bring to light some important properties of this phenomena in contexts more mathematically treatable.

\section{MODEL}

From the point of view of confinement as a violation of positivity of the two-point function, it is possible to define a scalar field model with a nonlocal mass term such that fundamental degrees of freedom are confined.
\vspace*{2pt}
\begin{eqnarray}\label{LAG}
{\cal L}_E = \frac{1}{2} \left[ \phi(x) \left( -\partial^2 + m^2 +\frac{\Lambda^4}{-\partial^2 + M^2} \right) \phi(x) \right] \, ,
\end{eqnarray}
\vspace*{2pt}
where $m$, $M$ and $\Lambda$ are mass parameters, (roughly in the $0.1$ GeV scale), so we say that the scalar field $\phi$ has a momentum dependent mass, given by
\begin{eqnarray}\label{MASSP}
{\cal M}^2 (p) = m^2 + \frac{\Lambda^4}{p^2 + M^2} \, .
\end{eqnarray}
Note that the Lagrangian (\ref{LAG}) can be seen as a (very) simplified version of refined Gribov-Zwanziger (RGZ) \cite{Dudal:2008sp} Lagrangian, where the $A_\mu$ gauge field in the adjoint representation of an $SU(N_c)$ color group has been replaced by a simple scalar field without internal color structure. Another important detail is that in this effective theory for low energy QCD, the nonlocal term in (\ref{LAG}) (which gives rise to the nontrivial mass function (\ref{MASSP})) corresponds to the horizon function of Gribov (or Gribov's horizon) \cite{Gribov:1977wm,Vandersickel:2012tz} in its quadratic approximation in the gauge fields. We can easily visualize this by making a simple comparison with the Zwanziger horizon function
%Observe que a Lagrangiana (\ref{LAG}) pode ser vista como uma versão (muito) simplificada da Lagrangiana de Gribov-Zwanziger rininada (RGZ), onde o campo de calibre $A_\mu$ na representação adjunta de um grupo $SU(N_c)$ foi substituido por um simples campo escalar sem estrutura interna de cor. Outro detalhe importante é que nesta teoria efetiva para a QCD em baixas energias, o termo não local em (\ref{LAG}) (que dá origem à função de massa não trivial (\ref{MASSP})) corresponde à função horizonte de Gribov  \cite{Gribov:1977wm,Vandersickel:2012tz}, em sua aproximação quadrática nos campos de calibre. Podemos visualizer isso facilmente ao fazer uma simples comparação com a função horizonte
%
\begin{eqnarray}
H(A) = \gamma^2 \int d^4 x \, g f^{bal} A_{\mu}^{a} (\mathpzc{M}^{-1})^{lm} g f^{bkm} A_{\mu}^{k} \, ,
\end{eqnarray}
where $\mathpzc{M}^{-1}$ is the nonlocal term, called the Faddeev-Popov operator, which is given in the Landau gauge by
\begin{eqnarray}
\mathpzc{M}^{ab} = -\partial^2 \delta^{ab} + g f^{abc} A_{\mu}^{c} \partial_\mu \, ,
\end{eqnarray}
and $\gamma$ is the Gribov parameter.

In the next section, we will look at the propagator of this theory and compare it with the RGZ propagator.

\section{FREE PROPAGATOR AT FINITE TEMPERATURE}

\begin{figure}[!ht]
	\centering
	\includegraphics[scale=2.2]{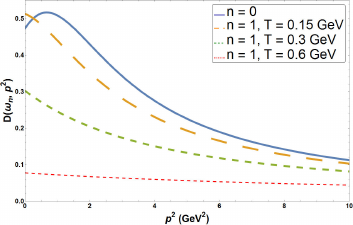}
	\caption{Free propagator of a scalar particle. The solid line corresponds to the first Matsubara mode (n = 0) which corresponds to T = 0. The dashed lines correspond to the second Matsubara mode (n = 1) for three different temperatures.}
	\label{fig1}
\end{figure}

First we are interested in obtaining the free propagator for such a model, so we can see if it is comparable with the RGZ propagator. Thus, the free propagator is given by
\begin{eqnarray}
D_{F}(\omega_n , \vec{p}\,{^2}) &=& \frac{p{^2} + M^2}{(p{^2} + M^2)(p{^2} + m^2) + \Lambda^4} \nonumber \\
\nonumber \\
%\vspace{30pt}
&=& \frac{\omega_n ^2 + \vec{p}\,{^2} + M^2}{(\omega_n ^2 + \vec{p}\,{^2} - \Lambda_1)(\omega_n ^2 + \vec{p}\,{^2} - \Lambda_2)} \, ,
\end{eqnarray}
where $\Lambda_1$ and $\Lambda_2$ are given by
\begin{eqnarray}
\Lambda_1 = \frac{- (m^2 + M^2) + \sqrt{(m^2 - M^2)^2 - 4\Lambda^4}}{2}
\end{eqnarray}
and
\begin{eqnarray}
\Lambda_2 = \frac{- (m^2 + M^2) - \sqrt{(m^2 - M^2)^2 - 4\Lambda^4}}{2} \, ,
\end{eqnarray}
and $\omega_n = 2\pi n T$ are the Matsubara frequencies at temperature T, with $n \in \mathbb{Z}$. Figure \ref{fig1} shows the behavior of the free propagator with respect to the momentum squared, for the zeroth and for the first Matsubara mode at different temperatures.
%
%\begin{minipage}{\linewidth}
%    \begin{center}
		%\captionof{figure}{Contorno $C$.} 
%		\label{fig1}
%		\includegraphics[scale=0.23]{graficoD.pdf}
		%\legend{Contorno da integral na eq. (60).}
		%\source{O autor}	
%	\end{center}
%\end{minipage}
%

\begin{figure}[!ht]
	\centering
	\includegraphics[scale=0.4]{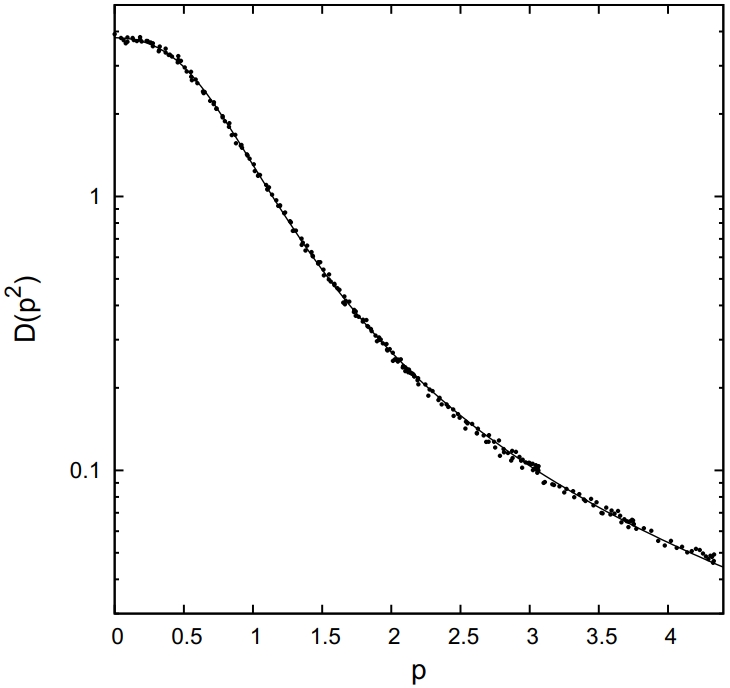}
	\caption{SU(2) gluon form factor for a lattice volume $V = 128^4$ \cite{Cucchieri:2012gb}.}
	\label{fig2}
\end{figure}

\noindent
We can see that for low momenta, the second derivative of the free propagator with respect to $p^2$ is negative, i.e., it is in a confining regime. However, by introducing temperature (Matsubara modes, in this case the first mode), as it increases, we see that such a derivative tends to zero. Therefore, of course, we may think that from a certain temperature value the free propagator will no longer have the positivity violation characteristic. As, in this work, such violation implies confinement, from this temperature value the particles would then be ``deconfined''. The values used, to have a notion of how is the behavior of the free propagator, for the mass parameters were \cite{Cucchieri:2012gb}: $M^2 = 2.51 \, \mbox{GeV}^2$, $m^2 = -1.91 \, \mbox{GeV}^2$ e $\Lambda^4 = 5.31 \, \mbox{GeV}^4$. We also see that for the Matsubara zero mode, the graph shows a curve very similar to that of the RGZ gluon form factor at T = 0, showing that this model, although simple, really resembles the RGZ model which in turn presents a propagator to the gluon qualitatively similar to the gluon propagator in the lattice, as we can see in the figure \ref{fig2}.
%Podemos perceber que para baixos momentos, a derivada segunda do propagador livre com relação ao momento é negativa, ou seja, regime confinante. Entretanto, ao introduzir a temperatura (os modos de Matsubara, neste caso o primeiro modo), comforme ela vai aumentando, observamos que tal derivada vai tendendo a zero para quaisquer valores de momento. Logo, naturalmente, podemos pensar que apartir de um certo valor de temperatura o propagador livre não apresentará mais a característica de violação de positividade. Como, neste trabalho, tal violção implica em confinamento, a partir deste valor de temperatura as partículos estariam então ``desconfinadas''. Os valores usados, para termos uma noção de como é o comportamento do propagador livre, para os parâmetros de massa foram \cite{Cucchieri:2012gb}: $M^2 = 2.51 \, GeV^2$, $m^2 = -1.91 \, GeV^2$ e $\Lambda^4 = 5.31 \, GeV^4$. Vemos também que para o modo zero de Matsubara, o gráfico apresenta uma curva bastante parecida com a da RGZ, mostrando então que este modelo, embora simples, realmente assemelha-se com o modelo da RGZ que por sua vez apresenta um propagador para o glúon muito próximo ao propagador do glúon na rede, como podemos ver na figura \ref{fig2}.
%
%\begin{minipage}{\linewidth}
%    \begin{center}
		%\captionof{figure}{Contorno $C$.} 
%		\label{fig1}
%		\includegraphics[scale=0.4]{graficoD2.jpg}
		%\legend{Contorno da integral na eq. (60).}
		%\source{O autor}	
%	\end{center}
%\end{minipage}
%

\noindent
In the next section we will obtain the free partition function, in order to investigate the behavior of the thermodynamic variables.

\section{FREE PARTITION FUNCTION}

\begin{figure}[!h]
	\centering
	\includegraphics[scale=2.2]{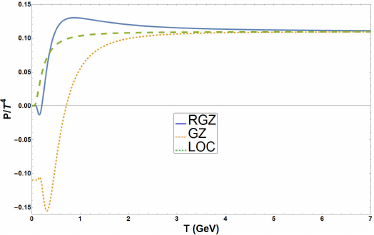}
	\caption{Behavior of pressure for a free gas. The solid line corresponds to the RGZ model, while the dashed lines corresponds to the similar GZ model and free massive gas model (LOC) (green and orange) respectively.}
	\label{fig3}
\end{figure}

\begin{figure}[!h]
	\centering
	\includegraphics[scale=2.2]{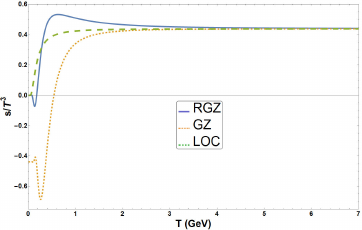}
	\caption{Behavior of the entropy density for a free gas. The solid line corresponds to the RGZ model, while the dashed lines corresponds to the similar GZ model and the free massive gas model (LOC) (green and orange) respectively.}
	\label{fig4}
\end{figure}

One of our objectives is to analyze the thermodynamic variables of this model. For this, we will use the free partition function, which is given by
\begin{equation}
\begin{aligned}
Z_{F}(\beta) &= \int {\cal D}\phi \; \mbox{exp}^{- S_E (\beta)} \\
&= \int{\cal D} \phi \; \mbox{exp} \left\lbrace - \frac{1}{2} \int d^4 x \right. \\
& \left. \hspace*{6pt} \times \; \phi(x) \left( -\partial^2 + m^2 +\frac{\Lambda^4}{-\partial^2 + M^2} \right) \phi(x) \right\rbrace \, .
\end{aligned}
\end{equation}
From it we can then obtain the free energy, which is given by
\begin{eqnarray}
\Omega = - \frac{1}{\beta} \; \mbox{ln} \, Z_{F}(\beta) \; ,
\end{eqnarray}
and consequently, obtain some thermodynamic variables for this simple model of a free gas, as for example the pressure. For a better analysis of this variable, figure \ref{fig3} shows a comparison between models. Recalling that our model resembles RGZ, we then compared with a similar model to that of Gribov-Zwanziger (GZ), which corresponds to $M = 0$ and $m = 0$ in (\ref{LAG}), and also with a model free massive gas with $m_{free} = m$ $(\Lambda = 0)$.
We can see that for high temperatures (energies), the models converge to the Stefan-Boltzmann limit. However for low temperatures, certain oscillations occur, implying a negative pressure. We see that, for our model, in the region where the pressure is negative, i.e. in the temperature range between (approximately) 0.07 and 0.2 GeV, the positivity violation still appears as we see from figure \ref{fig1}.
%Vemos que, para o nosso modelo, na região onde a pressão é negativa, ou seja, no intervalo de temperatura entre (aproximadamente) 0.07 à 0.2 GeV, a violação de positividade ainda aparece ao olharmos para a figura \ref{fig1}.
Another important variable to be studied is the entropy density. The graph figure \ref{fig4}, analogous to pressure, makes a comparison with other models as well (RGZ, GZ and model with a local Lagrangian i.e., a free massive gas).

\noindent
We also perceive an agreement of the models for high energies, but for low energies, there are again certain oscillations and negative entropy.

So far we have seen that the model suggested for free scalar field is comparable with RGZ, even though it is a simple model (toy model). Our future goal is then to take into account the interaction and analyze its influence on the thermodynamic variables.

\section{CONCLUSIONS AND PERSPECTIVES}

The RGZ propagator has great agreement with the calculation of the lattice, but when one introduces the temperature in the theory some inconsistencies appear that are the same presented here for this simple model of scalar fields. Since QCD is a theory that describes the strong interactions, considering the interaction terms is obviously something important in this model, so one of the future calculations will be to obtain the first order correction to the propagator, and consequently for the partition function and for the thermodynamic variables in order to, perhaps, see some improvement in the results.

\section{\label{sec:level7}ACKNOWLEDGMENTS}

A. V. Silva would to thank the organizers for the rich discussion environment at Hadron Physics 2018, where this poster was presented. This work was partially financed by the Brazilian funding agencies CAPES and FAPERJ.

\bibliographystyle{apsrev4-1}
\bibliography{refs}

\begin{thebibliography}{23}%
\makeatletter
\providecommand \@ifxundefined [1]{%
 \@ifx{#1\undefined}
}%
\providecommand \@ifnum [1]{%
 \ifnum #1\expandafter \@firstoftwo
 \else \expandafter \@secondoftwo
 \fi
}%
\providecommand \@ifx [1]{%
 \ifx #1\expandafter \@firstoftwo
 \else \expandafter \@secondoftwo
 \fi
}%
\providecommand \natexlab [1]{#1}%
\providecommand \enquote  [1]{``#1''}%
\providecommand \bibnamefont  [1]{#1}%
\providecommand \bibfnamefont [1]{#1}%
\providecommand \citenamefont [1]{#1}%
\providecommand \href@noop [0]{\@secondoftwo}%
\providecommand \href [0]{\begingroup \@sanitize@url \@href}%
\providecommand \@href[1]{\@@startlink{#1}\@@href}%
\providecommand \@@href[1]{\endgroup#1\@@endlink}%
\providecommand \@sanitize@url [0]{\catcode `\\12\catcode `\$12\catcode
  `\&12\catcode `\#12\catcode `\^12\catcode `\_12\catcode `\%12\relax}%
\providecommand \@@startlink[1]{}%
\providecommand \@@endlink[0]{}%
\providecommand \url  [0]{\begingroup\@sanitize@url \@url }%
\providecommand \@url [1]{\endgroup\@href {#1}{\urlprefix }}%
\providecommand \urlprefix  [0]{URL }%
\providecommand \Eprint [0]{\href }%
\providecommand \doibase [0]{http://dx.doi.org/}%
\providecommand \selectlanguage [0]{\@gobble}%
\providecommand \bibinfo  [0]{\@secondoftwo}%
\providecommand \bibfield  [0]{\@secondoftwo}%
\providecommand \translation [1]{[#1]}%
\providecommand \BibitemOpen [0]{}%
\providecommand \bibitemStop [0]{}%
\providecommand \bibitemNoStop [0]{.\EOS\space}%
\providecommand \EOS [0]{\spacefactor3000\relax}%
\providecommand \BibitemShut  [1]{\csname bibitem#1\endcsname}%
\let\auto@bib@innerbib\@empty
%</preamble>
\bibitem [{\citenamefont {Cornwall}(1980)}]{5Cornwall:1980zw}%
  \BibitemOpen
  \bibfield  {author} {\bibinfo {author} {\bibfnamefont {J.~M.}\ \bibnamefont
  {Cornwall}},\ }\href {\doibase 10.1103/PhysRevD.22.1452} {\bibfield
  {journal} {\bibinfo  {journal} {Phys. Rev. D}\ }\textbf {\bibinfo {volume}
  {22}},\ \bibinfo {pages} {1452} (\bibinfo {year} {1980})}\BibitemShut
  {NoStop}%
%%CITATION = PHRVA,D22,1452;%%
\bibitem [{\citenamefont {Brambilla}\ \emph {et~al.}(2014)\citenamefont
  {Brambilla} \emph {et~al.}}]{6Brambilla:2014jmp}%
  \BibitemOpen
  \bibfield  {author} {\bibinfo {author} {\bibfnamefont {N.}~\bibnamefont
  {Brambilla}} \emph {et~al.},\ }\href {\doibase
  10.1140/epjc/s10052-014-2981-5} {\bibfield  {journal} {\bibinfo  {journal}
  {Eur. Phys. J. C}\ }\textbf {\bibinfo {volume} {74}},\ \bibinfo {pages}
  {2981} (\bibinfo {year} {2014})},\ \Eprint {http://arxiv.org/abs/1404.3723}
  {arXiv:1404.3723 [hep-ph]} \BibitemShut {NoStop}%
%%CITATION = ARXIV:1404.3723;%%
\bibitem [{\citenamefont {Alkofer}\ and\ \citenamefont {von
  Smekal}(2001)}]{7Alkofer:2000wg}%
  \BibitemOpen
  \bibfield  {author} {\bibinfo {author} {\bibfnamefont {R.}~\bibnamefont
  {Alkofer}}\ and\ \bibinfo {author} {\bibfnamefont {L.}~\bibnamefont {von
  Smekal}},\ }\href {\doibase 10.1016/S0370-1573(01)00010-2} {\bibfield
  {journal} {\bibinfo  {journal} {Phys. Rept.}\ }\textbf {\bibinfo {volume}
  {353}},\ \bibinfo {pages} {281} (\bibinfo {year} {2001})},\ \Eprint
  {http://arxiv.org/abs/hep-ph/0007355} {arXiv:hep-ph/0007355 [hep-ph]}
  \BibitemShut {NoStop}%
%%CITATION = HEP-PH/0007355;%%
\bibitem [{\citenamefont {Aguilar}\ and\ \citenamefont
  {Natale}(2004)}]{8Aguilar:2004sw}%
  \BibitemOpen
  \bibfield  {author} {\bibinfo {author} {\bibfnamefont {A.~C.}\ \bibnamefont
  {Aguilar}}\ and\ \bibinfo {author} {\bibfnamefont {A.~A.}\ \bibnamefont
  {Natale}},\ }\href {\doibase 10.1088/1126-6708/2004/08/057} {\bibfield
  {journal} {\bibinfo  {journal} {JHEP}\ }\textbf {\bibinfo {volume} {08}},\
  \bibinfo {pages} {057} (\bibinfo {year} {2004})},\ \Eprint
  {http://arxiv.org/abs/hep-ph/0408254} {arXiv:hep-ph/0408254 [hep-ph]}
  \BibitemShut {NoStop}%
%%CITATION = HEP-PH/0408254;%%
\bibitem [{\citenamefont {Aguilar}\ \emph {et~al.}(2008)\citenamefont
  {Aguilar}, \citenamefont {Binosi},\ and\ \citenamefont
  {Papavassiliou}}]{8Aguilar:2008xm}%
  \BibitemOpen
  \bibfield  {author} {\bibinfo {author} {\bibfnamefont {A.~C.}\ \bibnamefont
  {Aguilar}}, \bibinfo {author} {\bibfnamefont {D.}~\bibnamefont {Binosi}}, \
  and\ \bibinfo {author} {\bibfnamefont {J.}~\bibnamefont {Papavassiliou}},\
  }\href {\doibase 10.1103/PhysRevD.78.025010} {\bibfield  {journal} {\bibinfo
  {journal} {Phys. Rev. D}\ }\textbf {\bibinfo {volume} {78}},\ \bibinfo
  {pages} {025010} (\bibinfo {year} {2008})},\ \Eprint
  {http://arxiv.org/abs/0802.1870} {arXiv:0802.1870 [hep-ph]} \BibitemShut
  {NoStop}%
%%CITATION = ARXIV:0802.1870;%%
\bibitem [{\citenamefont {Maas}(2013)}]{9Maas:2011se}%
  \BibitemOpen
  \bibfield  {author} {\bibinfo {author} {\bibfnamefont {A.}~\bibnamefont
  {Maas}},\ }\href {\doibase 10.1016/j.physrep.2012.11.002} {\bibfield
  {journal} {\bibinfo  {journal} {Phys. Rept.}\ }\textbf {\bibinfo {volume}
  {524}},\ \bibinfo {pages} {203} (\bibinfo {year} {2013})},\ \Eprint
  {http://arxiv.org/abs/1106.3942} {arXiv:1106.3942 [hep-ph]} \BibitemShut
  {NoStop}%
%%CITATION = ARXIV:1106.3942;%%
\bibitem [{\citenamefont {Berges}\ \emph {et~al.}(2002)\citenamefont {Berges},
  \citenamefont {Tetradis},\ and\ \citenamefont {Wetterich}}]{10Berges:2000ew}%
  \BibitemOpen
  \bibfield  {author} {\bibinfo {author} {\bibfnamefont {J.}~\bibnamefont
  {Berges}}, \bibinfo {author} {\bibfnamefont {N.}~\bibnamefont {Tetradis}}, \
  and\ \bibinfo {author} {\bibfnamefont {C.}~\bibnamefont {Wetterich}},\ }\href
  {\doibase 10.1016/S0370-1573(01)00098-9} {\bibfield  {journal} {\bibinfo
  {journal} {Phys. Rept.}\ }\textbf {\bibinfo {volume} {363}},\ \bibinfo
  {pages} {223} (\bibinfo {year} {2002})},\ \Eprint
  {http://arxiv.org/abs/hep-ph/0005122} {arXiv:hep-ph/0005122 [hep-ph]}
  \BibitemShut {NoStop}%
%%CITATION = HEP-PH/0005122;%%
\bibitem [{\citenamefont {Blaizot}()}]{11blaizotl}%
  \BibitemOpen
  \bibfield  {author} {\bibinfo {author} {\bibfnamefont {J.-P.}\ \bibnamefont
  {Blaizot}},\ }\href@noop {} {\ }\Eprint {http://arxiv.org/abs/0801.0009}
  {arXiv:0801.0009 [cond-mat.stat-mech]} \BibitemShut {NoStop}%
%%CITATION = ARXIV:0801.0009;%%
\bibitem [{\citenamefont {Cloet}\ and\ \citenamefont
  {Roberts}(2014)}]{12Cloet:2013jya}%
  \BibitemOpen
  \bibfield  {author} {\bibinfo {author} {\bibfnamefont {I.~C.}\ \bibnamefont
  {Cloet}}\ and\ \bibinfo {author} {\bibfnamefont {C.~D.}\ \bibnamefont
  {Roberts}},\ }\href {\doibase 10.1016/j.ppnp.2014.02.001} {\bibfield
  {journal} {\bibinfo  {journal} {Prog. Part. Nucl. Phys.}\ }\textbf {\bibinfo
  {volume} {77}},\ \bibinfo {pages} {1} (\bibinfo {year} {2014})},\ \Eprint
  {http://arxiv.org/abs/1310.2651} {arXiv:1310.2651 [nucl-th]} \BibitemShut
  {NoStop}%
%%CITATION = ARXIV:1310.2651;%%
\bibitem [{\citenamefont {Karsch}(2009)}]{13Karsch:2009zz}%
  \BibitemOpen
  \bibfield  {author} {\bibinfo {author} {\bibfnamefont {F.}~\bibnamefont
  {Karsch}},\ }\href {\doibase 10.1016/j.ppnp.2008.12.024} {\bibfield
  {journal} {\bibinfo  {journal} {Prog. Part. Nucl. Phys.}\ }\textbf {\bibinfo
  {volume} {62}},\ \bibinfo {pages} {503} (\bibinfo {year} {2009})}\BibitemShut
  {NoStop}%
%%CITATION = PPNPD,62,503;%%
\bibitem [{\citenamefont {Brodsky}\ \emph {et~al.}(2015)\citenamefont
  {Brodsky}, \citenamefont {de~Teramond}, \citenamefont {Dosch},\ and\
  \citenamefont {Erlich}}]{14Brodsky:2014yha}%
  \BibitemOpen
  \bibfield  {author} {\bibinfo {author} {\bibfnamefont {S.~J.}\ \bibnamefont
  {Brodsky}}, \bibinfo {author} {\bibfnamefont {G.~F.}\ \bibnamefont
  {de~Teramond}}, \bibinfo {author} {\bibfnamefont {H.~G.}\ \bibnamefont
  {Dosch}}, \ and\ \bibinfo {author} {\bibfnamefont {J.}~\bibnamefont
  {Erlich}},\ }\href {\doibase 10.1016/j.physrep.2015.05.001} {\bibfield
  {journal} {\bibinfo  {journal} {Phys. Rept.}\ }\textbf {\bibinfo {volume}
  {584}},\ \bibinfo {pages} {1} (\bibinfo {year} {2015})},\ \Eprint
  {http://arxiv.org/abs/1407.8131} {arXiv:1407.8131 [hep-ph]} \BibitemShut
  {NoStop}%
%%CITATION = ARXIV:1407.8131;%%
\bibitem [{\citenamefont {Nambu}\ and\ \citenamefont
  {Jona-Lasinio}(1961)}]{3Nambu:1961tp}%
  \BibitemOpen
  \bibfield  {author} {\bibinfo {author} {\bibfnamefont {Y.}~\bibnamefont
  {Nambu}}\ and\ \bibinfo {author} {\bibfnamefont {G.}~\bibnamefont
  {Jona-Lasinio}},\ }\href {\doibase 10.1103/PhysRev.122.345} {\bibfield
  {journal} {\bibinfo  {journal} {Phys. Rev.}\ }\textbf {\bibinfo {volume}
  {122}},\ \bibinfo {pages} {345} (\bibinfo {year} {1961})},\ \bibinfo {note}
  {[,127(1961)]}\BibitemShut {NoStop}%
%%CITATION = PHRVA,122,345;%%
\bibitem [{\citenamefont {Klevansky}(1992)}]{15Klevansky:1992qe}%
  \BibitemOpen
  \bibfield  {author} {\bibinfo {author} {\bibfnamefont {S.~P.}\ \bibnamefont
  {Klevansky}},\ }\href {\doibase 10.1103/RevModPhys.64.649} {\bibfield
  {journal} {\bibinfo  {journal} {Rev. Mod. Phys.}\ }\textbf {\bibinfo {volume}
  {64}},\ \bibinfo {pages} {649} (\bibinfo {year} {1992})}\BibitemShut
  {NoStop}%
%%CITATION = RMPHA,64,649;%%
\bibitem [{\citenamefont {Levy}(1967)}]{16levy}%
  \BibitemOpen
  \bibfield  {author} {\bibinfo {author} {\bibfnamefont {M.}~\bibnamefont
  {Levy}},\ }\href@noop {} {\bibfield  {journal} {\bibinfo  {journal} {Nuovo
  Cim.}\ }\textbf {\bibinfo {volume} {52}},\ \bibinfo {pages} {23} (\bibinfo
  {year} {1967})}\BibitemShut {NoStop}%
\bibitem [{\citenamefont {Pisarski}\ and\ \citenamefont
  {Wilczek}(1984)}]{16Pisarski:1983ms}%
  \BibitemOpen
  \bibfield  {author} {\bibinfo {author} {\bibfnamefont {R.~D.}\ \bibnamefont
  {Pisarski}}\ and\ \bibinfo {author} {\bibfnamefont {F.}~\bibnamefont
  {Wilczek}},\ }\href {\doibase 10.1103/PhysRevD.29.338} {\bibfield  {journal}
  {\bibinfo  {journal} {Phys. Rev. D}\ }\textbf {\bibinfo {volume} {29}},\
  \bibinfo {pages} {338} (\bibinfo {year} {1984})}\BibitemShut {NoStop}%
%%CITATION = PHRVA,D29,338;%%
\bibitem [{\citenamefont {Rosenzweig}\ \emph {et~al.}(1980)\citenamefont
  {Rosenzweig}, \citenamefont {Schechter},\ and\ \citenamefont
  {Trahern}}]{16Rosenzweig:1979ay}%
  \BibitemOpen
  \bibfield  {author} {\bibinfo {author} {\bibfnamefont {C.}~\bibnamefont
  {Rosenzweig}}, \bibinfo {author} {\bibfnamefont {J.}~\bibnamefont
  {Schechter}}, \ and\ \bibinfo {author} {\bibfnamefont {C.~G.}\ \bibnamefont
  {Trahern}},\ }\href {\doibase 10.1103/PhysRevD.21.3388} {\bibfield  {journal}
  {\bibinfo  {journal} {Phys. Rev. D}\ }\textbf {\bibinfo {volume} {21}},\
  \bibinfo {pages} {3388} (\bibinfo {year} {1980})},\ \bibinfo {note}
  {[,3388(1979)]}\BibitemShut {NoStop}%
%%CITATION = PHRVA,D21,3388;%%
\bibitem [{\citenamefont {Lenaghan}\ \emph {et~al.}(2000)\citenamefont
  {Lenaghan}, \citenamefont {Rischke},\ and\ \citenamefont
  {Schaffner-Bielich}}]{16Lenaghan:2000ey}%
  \BibitemOpen
  \bibfield  {author} {\bibinfo {author} {\bibfnamefont {J.~T.}\ \bibnamefont
  {Lenaghan}}, \bibinfo {author} {\bibfnamefont {D.~H.}\ \bibnamefont
  {Rischke}}, \ and\ \bibinfo {author} {\bibfnamefont {J.}~\bibnamefont
  {Schaffner-Bielich}},\ }\href {\doibase 10.1103/PhysRevD.62.085008}
  {\bibfield  {journal} {\bibinfo  {journal} {Phys. Rev. D}\ }\textbf {\bibinfo
  {volume} {62}},\ \bibinfo {pages} {085008} (\bibinfo {year} {2000})},\
  \Eprint {http://arxiv.org/abs/nucl-th/0004006} {arXiv:nucl-th/0004006
  [nucl-th]} \BibitemShut {NoStop}%
%%CITATION = NUCL-TH/0004006;%%
\bibitem [{\citenamefont {Scavenius}\ \emph {et~al.}(2001)\citenamefont
  {Scavenius}, \citenamefont {Mocsy}, \citenamefont {Mishustin},\ and\
  \citenamefont {Rischke}}]{16Scavenius:2000qd}%
  \BibitemOpen
  \bibfield  {author} {\bibinfo {author} {\bibfnamefont {O.}~\bibnamefont
  {Scavenius}}, \bibinfo {author} {\bibfnamefont {A.}~\bibnamefont {Mocsy}},
  \bibinfo {author} {\bibfnamefont {I.~N.}\ \bibnamefont {Mishustin}}, \ and\
  \bibinfo {author} {\bibfnamefont {D.~H.}\ \bibnamefont {Rischke}},\ }\href
  {\doibase 10.1103/PhysRevC.64.045202} {\bibfield  {journal} {\bibinfo
  {journal} {Phys. Rev. C}\ }\textbf {\bibinfo {volume} {64}},\ \bibinfo
  {pages} {045202} (\bibinfo {year} {2001})},\ \Eprint
  {http://arxiv.org/abs/nucl-th/0007030} {arXiv:nucl-th/0007030 [nucl-th]}
  \BibitemShut {NoStop}%
%%CITATION = NUCL-TH/0007030;%%
\bibitem [{\citenamefont {Roder}\ \emph {et~al.}(2003)\citenamefont {Roder},
  \citenamefont {Ruppert},\ and\ \citenamefont {Rischke}}]{16Roder:2003uz}%
  \BibitemOpen
  \bibfield  {author} {\bibinfo {author} {\bibfnamefont {D.}~\bibnamefont
  {Roder}}, \bibinfo {author} {\bibfnamefont {J.}~\bibnamefont {Ruppert}}, \
  and\ \bibinfo {author} {\bibfnamefont {D.~H.}\ \bibnamefont {Rischke}},\
  }\href {\doibase 10.1103/PhysRevD.68.016003} {\bibfield  {journal} {\bibinfo
  {journal} {Phys. Rev. D}\ }\textbf {\bibinfo {volume} {68}},\ \bibinfo
  {pages} {016003} (\bibinfo {year} {2003})},\ \Eprint
  {http://arxiv.org/abs/nucl-th/0301085} {arXiv:nucl-th/0301085 [nucl-th]}
  \BibitemShut {NoStop}%
%%CITATION = NUCL-TH/0301085;%%
\bibitem [{\citenamefont {Dudal}\ \emph {et~al.}(2008)\citenamefont {Dudal},
  \citenamefont {Gracey}, \citenamefont {Sorella}, \citenamefont
  {Vandersickel},\ and\ \citenamefont {Verschelde}}]{Dudal:2008sp}%
  \BibitemOpen
  \bibfield  {author} {\bibinfo {author} {\bibfnamefont {D.}~\bibnamefont
  {Dudal}}, \bibinfo {author} {\bibfnamefont {J.~A.}\ \bibnamefont {Gracey}},
  \bibinfo {author} {\bibfnamefont {S.~P.}\ \bibnamefont {Sorella}}, \bibinfo
  {author} {\bibfnamefont {N.}~\bibnamefont {Vandersickel}}, \ and\ \bibinfo
  {author} {\bibfnamefont {H.}~\bibnamefont {Verschelde}},\ }\href {\doibase
  10.1103/PhysRevD.78.065047} {\bibfield  {journal} {\bibinfo  {journal} {Phys.
  Rev. D}\ }\textbf {\bibinfo {volume} {78}},\ \bibinfo {pages} {065047}
  (\bibinfo {year} {2008})},\ \Eprint {http://arxiv.org/abs/0806.4348}
  {arXiv:0806.4348 [hep-th]} \BibitemShut {NoStop}%
%%CITATION = ARXIV:0806.4348;%%
\bibitem [{\citenamefont {Gribov}(1978)}]{Gribov:1977wm}%
  \BibitemOpen
  \bibfield  {author} {\bibinfo {author} {\bibfnamefont {V.~N.}\ \bibnamefont
  {Gribov}},\ }\href {\doibase 10.1016/0550-3213(78)90175-X} {\bibfield
  {journal} {\bibinfo  {journal} {Nucl. Phys. B}\ }\textbf {\bibinfo {volume}
  {139}},\ \bibinfo {pages} {1} (\bibinfo {year} {1978})},\ \bibinfo {note}
  {[,1(1977)]}\BibitemShut {NoStop}%
%%CITATION = NUPHA,B139,1;%%
\bibitem [{\citenamefont {Vandersickel}\ and\ \citenamefont
  {Zwanziger}(2012)}]{Vandersickel:2012tz}%
  \BibitemOpen
  \bibfield  {author} {\bibinfo {author} {\bibfnamefont {N.}~\bibnamefont
  {Vandersickel}}\ and\ \bibinfo {author} {\bibfnamefont {D.}~\bibnamefont
  {Zwanziger}},\ }\href {\doibase 10.1016/j.physrep.2012.07.003} {\bibfield
  {journal} {\bibinfo  {journal} {Phys. Rept.}\ }\textbf {\bibinfo {volume}
  {520}},\ \bibinfo {pages} {175} (\bibinfo {year} {2012})},\ \Eprint
  {http://arxiv.org/abs/1202.1491} {arXiv:1202.1491 [hep-th]} \BibitemShut
  {NoStop}%
%%CITATION = ARXIV:1202.1491;%%
\bibitem [{\citenamefont {Cucchieri}\ \emph {et~al.}(2012)\citenamefont
  {Cucchieri}, \citenamefont {Dudal}, \citenamefont {Mendes},\ and\
  \citenamefont {Vandersickel}}]{Cucchieri:2012gb}%
  \BibitemOpen
  \bibfield  {author} {\bibinfo {author} {\bibfnamefont {A.}~\bibnamefont
  {Cucchieri}}, \bibinfo {author} {\bibfnamefont {D.}~\bibnamefont {Dudal}},
  \bibinfo {author} {\bibfnamefont {T.}~\bibnamefont {Mendes}}, \ and\ \bibinfo
  {author} {\bibfnamefont {N.}~\bibnamefont {Vandersickel}},\ }\href@noop {} {\
   (\bibinfo {year} {2012})},\ \Eprint {http://arxiv.org/abs/1202.0639}
  {arXiv:1202.0639 [hep-lat]} \BibitemShut {NoStop}%
%%CITATION = ARXIV:1202.0639;%%
\end{thebibliography}%

\end{document}